\documentclass[10pt, conference, a4paper]{IEEEtran}

\usepackage{booktabs} 

\usepackage{graphicx}
\usepackage{multirow}
\usepackage{cite}
\usepackage{array}
\usepackage{color}
\usepackage{amssymb}
\usepackage{amsmath}
\usepackage{amsthm}
\usepackage{enumitem}

\ifCLASSINFOpdf
\else
\fi
\newtheorem{theorem}{Theorem}
\newtheorem{definition}{Definition}

\newtheorem{lemma}{Lemma}

\newtheorem{observation}{Observation}

\newcommand{\com}[1]{} 

\def\eq{\triangleq}

\def\G{\mathcal{G}}
\def\N{\mathcal{N}}     
\def\S{\mathcal{S}}     
\def\W{W}               
\def\u{u}               
\def\ux{\widetilde{\u}} 
\def\s{s}               
\def\bs{\boldsymbol{\s}} 

\def\ex{\delta}   
\def\r{\rho}   
\def\bex{\boldsymbol{\ex}}   
\def\br{\boldsymbol{\r}}   

\def\bg{\beta}   

\def\tax{\phi}   
\def\rev{\pi}    

\begin{document}
%
\title{Achieving an Efficient and Fair Equilibrium Through Taxation \vspace{-3mm}
}

\author{
\IEEEauthorblockN{Lin Gao}
\IEEEauthorblockA{
Harbin Institute of Technology, Shenzhen\\
Email: gaol@hit.edu.cn\vspace{-10mm}
}
\and
\IEEEauthorblockN{Jianwei Huang}
\IEEEauthorblockA{
The Chinese University of Hong Kong\\
Email: jwhuang@ie.cuhk.edu.hk 
\vspace{-10mm}
}
\IEEEcompsocitemizethanks{
\IEEEcompsocthanksitem
This work is supported by the General Research Funds (Project Number CUHK 14206315 and CUHK 14219016) established under the University Grant Committee of the Hong Kong Special Administrative Region, China.
}}

\IEEEoverridecommandlockouts


%


\maketitle

\begin{abstract}
It is well known that a game  equilibrium  can be far from efficient or fair, due to the misalignment between individual and social objectives.
The focus of this paper is to design a new mechanism framework that induces an efficient and fair equilibrium in a general class of games.
To achieve this goal, we propose a \emph{taxation} framework, which first imposes a tax on each player based on the perceived payoff (income), and then redistributes the collected tax to other players properly.
By turning the tax rate, this framework spans the continuum space between strategic interactions (of selfish players) and altruistic interactions (of unselfish players), hence provides rich modelling possibilities.
The key challenge in the design of this framework is the proper taxing rule (i.e., the tax exemption and tax rate) that induces the desired equilibrium in a wide range of games.
First, we propose a \emph{flat} tax rate (i.e., a single tax rate for all players), which is necessary and sufficient for achieving an efficient equilibrium in \emph{any} static strategic game with common knowledge.
Then, we provide several tax exemption rules that achieve some typical fairness criterions (such as Nash bargaining solution and Shapley value) at the equilibrium.
We further discuss the incentive issue in the implementation of the proposed taxation mechanism, 
and illustrate the implementation of the framework in the game of Prisoners' Dilemma.
\end{abstract}


%


\section{Introduction}\label{sec:introduction}

\subsection{Background and Motivations}

A static strategic game models the situation where a set of rational (self-interested) players make interdependent choices simultaneously  \cite{game-theory1994}: each player's payoff depends not only on his own choice but also on the other players' choices.
An \emph{equilibrium} of a static strategic game   refers to a stable outcome of the game, i.e., a combination of players' \emph{preferred} choices from which none of them has the incentive to deviate.
Namely, at an equilibrium, no player can increase his payoff by unilaterally changing his choice, hence all players will stick on their choices at the equilibrium, leading to a stable outcome of the game.
Nowadays, {static} strategic game and its most well-known solution concept, \emph{Nash Equilibrium (NE)} \cite{game-Nash}, have been widely adopted for the modeling and analysis of competition and cooperation in various networking systems \cite{game-book2012,game-book2013}.

In many static strategic games, however, the game equilibrium is often \emph{not} efficient, due to the misalignment between the individual players' objectives and the social objective.
Specifically, a game outcome is said to be \emph{efficient}, or more precisely, socially efficient \cite{Pareto-2008}, if it maximizes the total payoff of all players (called the \emph{social welfare}).
Such an outcome is often called a \emph{Social Efficiency (SE)}.
Many existing works have investigated the performance (efficiency) gap between an NE and the SE of a game, where the notion of {Price-of-Anarchy (PoA)} \cite{PoA-2008} (resp.~{Price-of-Stability, PoS} \cite{PoS-2008}) was introduced to measures the social welfare ratio between the SE and the \emph{worst} NE (resp.~the \emph{best} NE).
 {However, both PoA and PoS are  \emph{descriptive} concepts, in the sense that they only characterize an existing situation (i.e., how good   an equilibrium is), without  providing any  constructive way to improve the situation.}


In this work, we aim to provide a \emph{constructive} way to improve the  efficiency of the static strategic game outcome.
Some recent works  have tried several ways based on the idea of ``\emph{Altruism}''
to induce the   efficient game equilibrium
\cite{altru-AGT2012,altru-TGC2010, altru-WINE2011, altru-EC2008, altru-INFOCOM2010, altru-ESA2009, altru-Xu2014, altru-CDC2010}.
The common idea of these works is {to implement an  \emph{Altruism Mechanism} on top of a game} to restructure the rules or payoffs of the game, so as to create an \emph{Altruistic} version of the game, wherein players are willing to behave in an altruistic manner.
More specifically, with the Altruism Mechanism, we modify each player's payoff  by adding a {positive component proportional to} the other players' payoffs \cite{altru-TGC2010,altru-Xu2014,altru-CDC2010} or the social welfare \cite{altru-AGT2012, altru-INFOCOM2010, altru-WINE2011, altru-EC2008}.
With such a modification, each player cares about not only his individual payoff (selfishness) but also the payoffs of others (altruism), hence is willing to behave in an altruistic or socially-aware manner.

However, the above Altruism Mechanism has several limitations.
First, it focuses only on the efficiency, without considering the \emph{fairness}.
Namely, it aims at improving the total payoff of all players (social welfare), while not considering the balance between the individual payoffs of different players.
Second, it often requires an additional \emph{budget} (i.e., the positive component 
added to each player's payoff) to incentivize the {altruistic} behaviors of players.
Sometimes such a budget can be huge and even infinite \cite{altru-AGT2012},\footnote{{In the mechanism of \cite{altru-AGT2012}, a requirement for an infinite budget implies that it is \emph{impossible} to achieve an efficient outcome in a specific game.}} which prevents the practical implementation of this mechanism.
 {Moreover, the Altruism Mechanism  is often designed for a specific set of static strategic games, and may not work in more general game settings.}
In this work, we aim to design such a mechanism that {achieves both \emph{efficiency} and \emph{fairness} in a wide range of static strategic games, together with the consideration of \emph{budget feasibility}.}




\vspace{-1mm}

\subsection{Solution Approach}
\vspace{-1mm}

We propose a novel \emph{taxation} framework to achieve our  goal.
{The key idea is to implement a properly designed \emph{Taxation Mechanism} on top of a game (resulting in a Taxation version of the game) to restructure the rules and payoffs of the game, such that the individual objectives are properly aligned with the social objective, hence reach the desired equilibrium.}
We design the Taxation Mechanism based on the idea of ``\emph{income tax} \cite{income-tax}'', which mainly consists of two parts:
\begin{enumerate}[leftmargin=6mm]
  \item \emph{\textbf{Taxing Rule}}, for imposing a \emph{tax} on each player according to the player's perceived payoff (income);
  \item  \emph{\textbf{Redistribution Rule}}, for allocating the collected tax (with budget deduction or compensation according to a pre-defined budget plan) to other players \emph{equally}.
\end{enumerate}

More specifically, the \emph{taxing rule} in our mechanism consists of two key components as in many real-world income tax systems \cite{income-tax}:
(i) a \emph{tax exemption} threshold for each player, denoting the minimal income level at which a player begins to pay positive taxes, and (ii) a \emph{tax rate} for each player, denoting the ratio at which a player is taxed (considering the impact of the tax exemption threshold).
{Intuitively, the above Taxation Mechanism redistributes the perceived payoffs of players, and makes players care about not only their individual payoffs (selfishness) but also the payoffs of others (altruism), hence reconciles the individual players'  objectives with the social objectives potentially.}

\begin{figure}
\vspace{-3mm}
\centering
\includegraphics[width=3.5in]{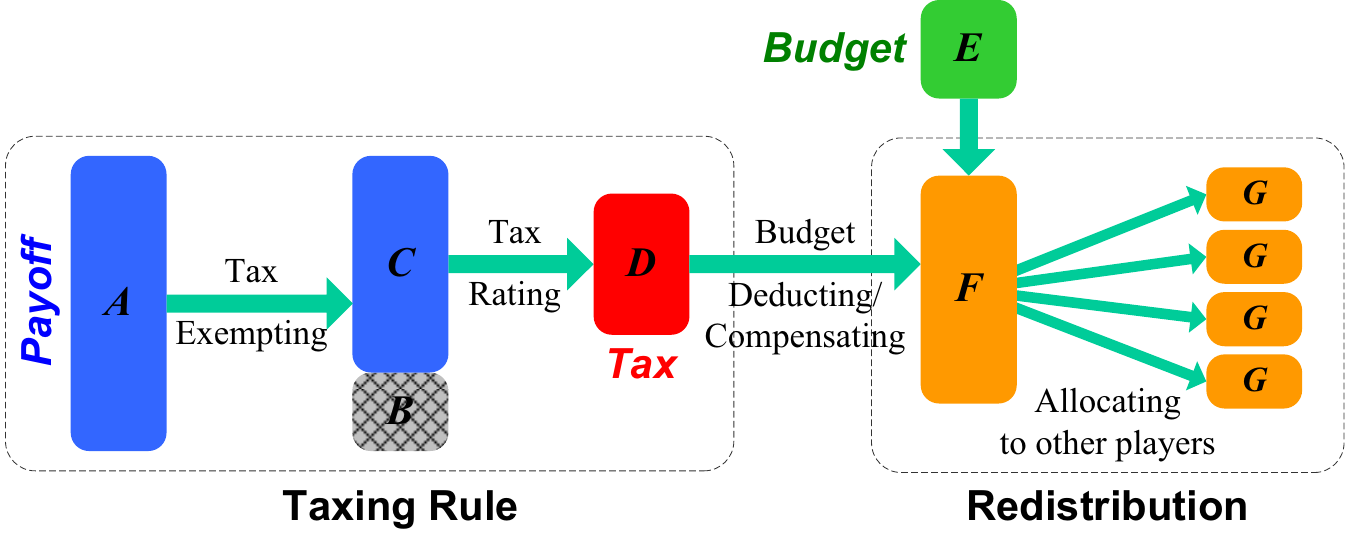}
\vspace{-6mm}
	\caption{The Taxation Mechanism.}
\label{fig:tax-model}
\vspace{-4mm}
\end{figure}

Figure \ref{fig:tax-model} illustrates our proposed mechanism, where $A$ is the total income (payoff) of a player, $B$ is the tax exemption, $C=A-B$ is the taxable income, $D$ is the   tax imposed on the player, $E$ is the pre-specified budget (which could be negative), $F=D+E$ is the income tax after budget deduction or compensation, which is then equally redistributed to other players (each getting a revenue of $G$).


Comparing with the   Altruism Mechanism in \cite{altru-AGT2012,altru-TGC2010, altru-WINE2011, altru-EC2008, altru-INFOCOM2010, altru-ESA2009, altru-Xu2014, altru-CDC2010},~this Taxation Mechanism has the following advantages.
First, it generalizes the existing Altruism Mechanism.
Second, it is more flexible due to the additional design dimension of tax exemption in the payoff reconstruction, which is critical for reaching the desirable fairness.
Third, it works with any given budget (even negative) from the platform.

\vspace{-1mm}

\subsection{Contributions}
\vspace{-1mm}

{We will study in details the   design of the Taxation Mechanism that induces the desired outcomes with the properties of efficiency and fairness, while considering the budget feasibility.
It is notable that the tax redistribution rule and the budget plan are given in advance, hence the key problem is \emph{to design a proper taxing rule (i.e., tax exemptions and tax rates for all players)}.}
Our key theoretical results are as follows:

\subsubsection{Efficiency}
We first show that the efficiency of outcome  only depends  on the tax rates, and is independent of the tax exemptions.
Based on this observation, we propose the efficient tax rates (that achieve the efficient outcome), which are identical (flat) for all players, and have a simple relationship with the player number and the budget.
{We further show that such an efficient flat tax rate is necessary and sufficient for achieving an efficient outcome in \emph{any} static strategic game with common knowledge \cite{common}.\footnote{{Common knowledge  is a mild assumption widely used in the area of game theory \cite{common}. It means that each player is fully aware of the rules and payoffs of the game, and meanwhile each player knows that each other player is fully aware of the rules and payoffs of the game.}}}

{It is worth noting that the above result doesn't rely on any additional assumption (except common knowledge).
The reason is that the above flat tax rate can guarantee that for any static strategic game (with common knowledge), its Taxation version must have at least one NE (which is the SE), even if the original game doesn't have an NE.}


\subsubsection{Fairness}
We show that the tax exemptions determine how the social welfare is distributed among players, given the efficient flat tax rate derived above.
Thus, by tuning the tax exemptions carefully, we can achieve any desired social welfare division, hence any fairness criterion.
As an illustrative example, we provide the tax exemptions that achieve several typical fairness criterions such as the Max-min fairness \cite{max-min}, in which the social welfare is divided (among players) in such a way that the minimal  payoff of players is maximized.  


\vspace{2mm}



To our best knowledge, this is the first work that proposes a   Taxation Mechanism for inducing both efficient and fair outcomes in static strategic games.
In addition to theoretical results mentioned above,
we also illustrate the implementation of the proposed mechanism in the \emph{Prisoners' Dilemma} game \cite{game-theory1994}, a widely-used example in game theory. 
The key contributions of this work are summarized  as follows.

\begin{itemize}[leftmargin=5mm]
  \item \emph{Novel Framework (Section \ref{sec:framework}):}
      The proposed Taxation Mechanism generalizes the existing Altruism Mechanism, and can induce the desired outcome with the properties of efficiency,  fairness, and budget feasibility, in \emph{any} static strategic game with common knowledge.
      Moreover, it spans the continuum space
        between strategic interactions (of selfish players) and altruistic interactions (of unselfish players), hence provides rich modeling possibilities.

  \item \emph{Efficiency and Fairness (Sections \ref{sec:framework-efficiency}--\ref{sec:framework-fairness}):}
        We propose a simple flat tax rate depending 
on the number of players and the budget, which is necessary and sufficient
for achieving an efficient outcome (in any static strategic game with
common knowledge). 
        We further show how to tune the tax exemptions to achieve a desirable fairness criterion. 

  \item \emph{Implementation and Incentive Issue (Section \ref{sec:framework-implement}):}
      We illustrate the implementation of the proposed mechanism in  the game of Prisoners' Dilemma, and discuss the related incentive issue in the practical implementation. 
\end{itemize}


\section{Related Work}\label{sec:literature}

The inefficiency of a game equilibrium has been long recognized in economics.
Many economists have investigated and measured such an efficiency loss, via the notions of PoA \cite{PoA-2008} and {PoS} \cite{PoS-2008}.
Some recent works   proposed several ways
to improve the efficiency of the game equilibrium
\cite{altru-AGT2012,altru-TGC2010, altru-WINE2011, altru-EC2008, altru-INFOCOM2010, altru-ESA2009, altru-Xu2014, altru-CDC2010}.
They implemented an ``Altruism Mechanism'' on top of a game, so as to alter each player's payoff by adding a positive component proportional to the other players' payoffs \cite{altru-TGC2010,altru-Xu2014,altru-CDC2010} or the social welfare \cite{altru-AGT2012, altru-INFOCOM2010, altru-WINE2011, altru-EC2008, altru-ESA2009}.
However, these works focused only on the efficiency in the outcome, without considering the \emph{fairness} issue.
Moreover, some of them (e.g., \cite{altru-AGT2012,altru-INFOCOM2010,altru-WINE2011, altru-TGC2010, altru-Xu2014, altru-ESA2009}) required an additional \emph{budget} to incentivize the {altruistic} behaviors of players.
Our proposed Taxation Mechanism generalizes the Altruism Mechanism, and
achieves both efficiency and fairness in a wide range of static strategic games with common knowledge, under any given (even negative) budget.
We summarize the key features of the existing works \cite{altru-AGT2012,altru-TGC2010, altru-WINE2011, altru-EC2008, altru-INFOCOM2010, altru-ESA2009, altru-Xu2014, altru-CDC2010} and our work in  Table~\ref{table:literature}.
 
 \begin{table}[t]
  \centering
  \caption{Summary of Existing Works}\label{table:literature}
\scalebox{0.88}[0.88]{
\begin{tabular}{|c|c|c|c|c|c|}
  \hline
    & \multirow{2}{*}{\textbf{Efficiency}}  & \multirow{2}{*}{\textbf{Fairness}} & \multicolumn{3}{|c|}{\textbf{Budget}}
    \\
  \cline{4-6}
    & & & Positive  &  Zero  &  Negative
    \\
  \hline
  \centering
  \cite{  altru-TGC2010, altru-Xu2014}  & $ \surd $ & $\times$  & $ \surd $ & $\times$ & $\times$
\\
  \hline
  \centering
  \cite{altru-EC2008, altru-CDC2010}  & $ \surd $ & $\times$ & $\times$ & $ \surd $ & $\times$
    \\
  \hline
  \centering
  \cite{altru-AGT2012, altru-INFOCOM2010,altru-WINE2011, altru-ESA2009}  & $ \surd $ & $\times$  & $ \surd $ & $\times$ & $\times$
    \\
      \hline
  \centering
   Our work & $ \surd $  & $ \surd $  & $ \surd $ & $ \surd $  &  $ \surd $
    \\
  \hline
\end{tabular}
}
\vspace{-3mm}
\end{table}


Although our proposed mechanism requires some assumptions (e.g., common knowledge), it has several encouraging features, comparing with the numerous existing mechanism design results.
First, it considers not only   efficiency but also   fairness at the   outcome.
Second, its implementation complexity is very low, while many mechanism design approaches involve solving problems that are  NP-hard.
Hence, our work opens up a new direction of game mechanism design under a different assumption from many existing results.


%
%
%

%

\section{The Taxation Framework}\label{sec:framework}

In this section, we provide the theoretical framework  of the proposed mechanism.
We first review some basic concepts in static strategic games,
then propose the Taxation Mechanism,
and finally derive the taxing rule (under any given budget) that achieves the efficiency and fairness in the outcome.

\subsection{Static Strategic Game}

We first review some basic concepts in static strategic games \cite{game-theory1994}, which are critical for our later discussions.




In a \emph{static} strategic game, a set~of~$N$ players make inter-dependent choices simultaneously.
Each player is \emph{rational} (self-interested), aiming at maximizing  his own payoff.
As in many existing game literature (e.g., \cite{game-theory1994}), we assume that players have \emph{common knowledge} of the game \cite{common}:
{each player knows the rules and payoffs
of the game,
and also knows that each other player knows the rules and payoffs of the game}.\footnote{Games without common knowledge require players to reveal their private information credibly, and is often the focus of incentive mechanism design (e.g., auction), which is beyond the scope of our work.}
Formally, a static strategic game can be defined as follows.


\begin{definition}[Static Strategic Game \cite{game-theory1994}]\label{def:game}
A static strategic game, denoted by  $\G \eq (\N, \{\S_i\}_{i\in\N}, \{\u_i\}_{i\in\N})$, is given~by:
\begin{itemize}[leftmargin=5mm]
  \item  \textbf{Player}: a set $\N = \{1,2, \cdots , N\}$ of rational players;
  \item  \textbf{Strategy}: a set $\S_i$ of strategies for each player~$i\in\N$;
  \item  \textbf{Payoff}: a payoff function $ \u_i:\S_1\times \cdots \times \S_N \rightarrow \mathbb{R}$ for each player $i\in\N$, which
    maps every possible strategy profile  to a real number, i.e., the payoff of player $i$.
\end{itemize}
\end{definition}

Let $\s_i \in \S_i$ denote the strategy of player $i\in\N$,
$\bs \eq (\s_1,\cdots,\s_N) \in \S_1\times \cdots \times \S_N $ denote the strategy profile of all players,
and $\bs_{-i} \eq (\s_1,\cdots,\s_{i-1},\s_{i+1},\cdots, \s_N)$ denote the strategy profile of all players other than $i$.
For notational convenience, we will  also write $\bs $ as  $ (\s_i, \bs_{-i})$.
Then, the \emph{payoff} of player $i$ under a particular strategy profile $\bs $ can be written as $\u_i(\bs)$ or $\u_i(\s_i, \bs_{-i})$.
The \emph{social welfare} is defined as the total payoff of all players, denoted by
\begin{equation}\label{eq:sw}
\textstyle
\W (\bs) = \sum\limits_{i\in\N} \u_i(\bs) .
\end{equation}

An NE refers to a stable strategy profile (outcome) from which none of the players has the incentive to deviate \cite{game-Nash}.

\begin{definition}[Nash Equilibrium---NE \cite{game-Nash}]\label{def:NE}
A strategy profile $\bs^* \eq (\s_1^*,\cdots,\s_N^*)$ is an NE, if for every player $i\in\N$,
$$
\u_i(\s_i^*, \bs_{-i}^*) \geq \u_i(\s_i', \bs_{-i}^*),\quad \forall \s_i' \in \S_i .
$$
\end{definition}

%
%
%


A strategy profile (outcome) is socially efficient \cite{Pareto-2008}, or simply efficient, if it maximizes the social welfare in defined \eqref{eq:sw}.
Such an outcome is often called a Social Efficiency (SE).

\begin{definition}[Social Efficiency---SE \cite{Pareto-2008}]\label{def:SE}
A strategy profile $\bs^\circ \eq (\s_1^\circ,\cdots,\s_N^\circ)$ is an SE, if
$$
W ( \bs^\circ ) \geq W( \bs' ) , \quad \forall \bs' \in \S_1\times \cdots \times \S_N.
$$
\end{definition}

Note that an NE describes what is likely to occur  in the game as a stable outcome, while an SE describes what is desirable from the social perspective.
In practice, however, there can be a large gap between NE and SE  \cite{PoA-2008, PoS-2008}.
Our focus in this work is to restructure the rules   of the game through a properly designed mechanism, such that the modified game produces an NE   that is efficient and fair for the original game, i.e., ``\emph{what is likely to happen (in the modified game) is what is desired (for the original game)}''.

\subsection{The Taxation Mechanism}

We propose a Taxation framework, which implements a \emph{Taxation Mechanism} on top of a game to restructure the rules or payoffs of the game and achieve the desired outcome via the modified game.
As shown in Figure \ref{fig:tax-model}, the {Taxation Mechanism} is designed based on the idea of ``\emph{income tax} \cite{income-tax}'', and  mainly consists of two parts: (i) a \emph{taxing rule} for imposing income tax on each player, and (ii) a \emph{redistribution rule} for reallocating the imposed   tax back to players.

\subsubsection{Taxing Rule}
As in many real-world income tax systems \cite{income-tax}, our  taxing rule   consists of two components:
\begin{itemize}[leftmargin=5mm]
  \item A \emph{tax exemption} $ \ex_i \geq 0 $ for each player $i \in \N $, denoting the minimum income level at which he begins to pay positive taxes;  
  \item A \emph{tax rate} $\r_i \in [0,1] $ for each player $i \in \N $, denoting the ratio at which he is taxed, considering his tax exemption.\footnote{Corresponding to Figure \ref{fig:tax-model}, we have: $\ex_i = B$ and $\r_i = \frac{D}{C} = \frac{D}{A-B} $.}
\end{itemize}
For convenience, we denote $\bex \eq (\ex_1,\cdots, \ex_N)$ as the tax exemption vector and $\br \eq (\r_1,\cdots, \r_N)$ as the tax rate vector.
Then, the taxing rule can be formally written as $\{\bex, \br\}$, which specifies a taxing rule $\{\ex_i, \r_i\}$   for every player $i\in\N$.

Given the taxing rule $\{\ex_i, \r_i\}$ for player $i $, the   income tax imposed on player $i$ under a strategy profile $\bs$ is:\footnote{Precisely, we should write the income tax $\tax_i (\bs)$ as $\tax_i (\bs, \ex_i, \r_i)$. Here we omit $\ex_i$ and $ \r_i$ mainly for the purpose of writing convenience.}
\begin{equation}\label{eq:income-tax}
\tax_i (\bs)  = ( \u_i (\bs) - \ex_i ) \cdot \r_i,
\end{equation}
where $ \u_i (\bs)  $ is the total income (payoff) of player $i$ under $\bs$, 
and $\u_i (\bs) - \ex_i$ is the taxable income.


\subsubsection{Redistribution Rule}
The   tax collected from a player is  first manipulated (deducted or compensated) according to a pre-defined budget plan (i.e., $E$ in Figure \ref{fig:tax-model}), and then allocated to other players \emph{equally}.
Let $\bg > 0 $ denote the ratio of the tax after manipulation and before manipulation, called \emph{budget factor}.
In Figure \ref{fig:tax-model}, we have: $\bg  = \frac{F}{D} = \frac{D+E}{D} $.
The budget factor  $\bg$   fully characterizes a budget plan:
\begin{itemize}[leftmargin=5mm]
  \item If $ \bg < 1$, then $F = \bg D  < D$, implying  that the platform extracts a portion  $(1 - \bg)D$ of   received tax as   revenue;
  \item If $\bg >1$, then $F = \bg D > D$, implying that the platform provides a budget $(\bg-1)D$  for players as subsidy;
  \item If $ \bg = 1$, then $F = D$, implying that that the platform neither extracts revenue nor provides subsidy, hence the system is   budget balance.
\end{itemize}

Given the budget factor $\bg$,
the revenue that each player (other than   $i$)  gains from the tax $\tax_i (\bs)$ of   player $i$ is,
\begin{equation}\label{eq:tax-return}
\textstyle
\rev_i (\bs)  =  \tax_i(\bs) \cdot \bg \cdot \frac{1}{N-1},
\end{equation}
where the factor $\frac{1}{N-1}$ implies that $N-1$ other players share the tax (multiply by the budget factor $\bg$) equally.

\subsubsection{Taxation Version of Game}

Based on the above taxing rule and redistribution rule, we can define the new payoff of each player $i \in \N $ under the taxation mechanism as:
\begin{equation}\label{eq:payoff-tax}
\textstyle
\ux_i (\bs)   = \u_i (\bs) - \tax_i (\bs) + \sum\limits_{j \neq i} \rev_j (\bs).
\end{equation}
We refer to the new game with the new payoffs defined  in \eqref{eq:payoff-tax} as a \emph{Taxation Version} of the original game. Formally,

\begin{definition}[Taxation Version of Game]\label{def:taxation-game}
For any static strategic game $\G = (\N, \{\S_i\}_{i\in\N}, \{\u_i\}_{i\in\N})$, a $(\bex, \br)$-Taxation Version of $\G$, denoted by $\G(\bex, \br)$, is defined as
\begin{equation}\label{eq:taxversion}
\G(\bex, \br) \eq (\N, \{\S_i\}_{i\in\N}, \{\ux_i\}_{i\in\N}),
\end{equation}
where $\ux_i $ is given in \eqref{eq:payoff-tax}, i.e., the new payoff of player $i$ in the Taxation Mechanism with the taxing rule $\{\bex, \br\}$.
\end{definition}

Intuitively, a Taxation Version of a game represents a joint system of the original game   and the Taxation Mechanism, where players first interact and get paid according to the original game rules, and then are taxed based on the perceived payoffs according to the Taxation Mechanism.

Our focus in this work is to design a proper taxing rule $\{\bex, \br\}$ such that the $(\bex, \br)$-Taxation Version of \emph{any} static strategic game $\G$ (with common knowledge) produces the desired outcome for the original game $\G$.
We emphasize that  the taxing rule $\{\bex, \br\}$ is what we need to design, while the budget plan (i.e., $\bg$) is a given system parameter.

Before designing the taxing rule, we first provide several useful properties for the above $(\bex, \br)$-Taxation Version.\footnote{Due to space limit, we put all the proofs  in the technical report \cite{report}.}


\begin{lemma}\label{lemma:1}
With a flat tax rate (i.e., $\r_i=\r , \forall i \in \N$),
an SE of $\G(\bex, \br)$ must be an SE of  $\G$, and vice versa.
\end{lemma}

{Lemma \ref{lemma:1} states that if a flat tax rate is adopted, then an efficient outcome of the Taxation Version game $\G(\bex, \br)$ is also efficient to the original game $\G$, and vice versa.
This inspires us to focus on those taxing rules with the \emph{flat tax rate}.
However, Lemma \ref{lemma:1} does not tell us how to reach the SE at an equilibrium of $\G(\bex, \br)$.
 We will further discuss this   in Theorem \ref{thm:efficiency} of the next section.

%
%
%

We further notice that with a particular flat tax rate $\r$, the   payoff of each player $i$ defined in \eqref{eq:payoff-tax} can be written as:
 \begin{equation} \label{eq:xxx}
 \textstyle
 \ux_i(\bs) = (1-\r)\cdot \u_i(\bs) +  \r \cdot a \cdot \sum\limits_{j\neq i}\u_j(\bs) + \r \cdot b,
 \end{equation}
where  $a \eq \frac{\bg}{N-1}$ and $b \eq \ex_i- $ $\frac{\bg}{N-1} \cdot \sum_{j\neq i}\ex_j $.
From \eqref{eq:xxx}, we can see that with the increase of $\r $, each player  cares  less about his own payoff (less selfish), but more  about other players' payoffs (more unselfish).
Therefore, \emph{our Taxation framework spans the continuum space between strategic  interactions (of selfish players) and altruistic interactions (of unselfish players) by turning the taxing rate $\r$ from $0$ to $1$}.


\section{Efficiency}\label{sec:framework-efficiency}

We now study the efficient taxing rule $\{\bex, \br\}$ that produces the efficient outcome.
Inspired by Lemma \ref{lemma:1}, we narrow our focus within the taxing rules with the flat tax rate (i.e., $\r_i = \r,\forall i\in \N$).
In this case, an efficient outcome for the Taxation Version of a game $\G$ is also efficient for  $\G$.

For notational convenience, we define
$c \eq \frac{\bg}{N-1+\bg}$ and  $\Delta \eq \sum_{i \in \N}\ex_i $,  both are constant with regard to $\bs$.
Formally, we have the following efficient taxing rule.

\begin{theorem}[\textbf{Efficiency}]\label{thm:efficiency}
A taxing rule $\{\bex, \br\}$ is efficient for any tax exemptions $\bex$ under 
  the following flat tax rate, 
 \begin{equation} \label{eq:flat-rate}
\r_i = \r = \frac{N-1}{N-1+\bg}, \quad \forall i\in\N.
 \end{equation}
\end{theorem}

With the above tax rule in \eqref{eq:flat-rate}, for any static game $\G$, its Taxation Version $\G (\bex, \br) $ will generate an efficient outcome that maximizes the social welfares of $\G (\bex, \br) $ in \eqref{eq:taxversion} and $\G$.
Theorem \ref{thm:efficiency} shows that the efficiency of a taxing rule $\{\bex, \br\}$ only depends on the tax rates $\br $, while is independent of the tax exemptions $\bex$.
Moreover, it shows that a flat tax rate \eqref{eq:flat-rate} is sufficient for achieving an efficient outcome for any static strategic game with common knowledge.

{It is worth noting that \emph{even if
the original game $\G$ does not have an equilibrium,
our framework with the proposed flat tax rate \eqref{eq:flat-rate} can still induce an efficient equilibrium outcome from the Taxation Version $\G (\bex, \br) $}. 
This is because for any static strategic game $\G$, its Taxation Version $\G (\bex, \br) $ must have at least one efficient equilibrium (i.e.,  the SE) with the flat tax rate \eqref{eq:flat-rate}.} 
In fact, the above Taxation Version $\G (\bex, \br) $ is a \emph{potential game}, with the social welfare in \eqref{eq:sw} as the potential function.

We will   show that the flat tax rate  \eqref{eq:flat-rate} is also the \emph{unique} tax rate that can achieve efficient outcomes in \emph{all} static
strategic games with common knowledge.
Formally,

\begin{theorem}[\textbf{Uniqueness}]\label{thrm:unique}
A taxing rule $\{\bex, \br\}$ is efficient for all static strategic games (with common knowledge), if and only if the tax rates satisfy \eqref{eq:flat-rate}.
\end{theorem}

We have shown the sufficiency of \eqref{eq:flat-rate} in Theorem \ref{thm:efficiency}, hence only need to prove the necessity of \eqref{eq:flat-rate}, that is, ``there does \emph{not} exist a different taxing rule $\{\bex, \br\}$ from \eqref{eq:flat-rate} that is efficient for \emph{all} static strategic games (with common knowledge)''.
To prove this, we only need to find a static strategic game in which the taxing rule $\{\bex, \br\}$ satisfying \eqref{eq:flat-rate} is the \emph{only} taxing rule that can generate efficient outcomes.


We can further see that the efficient tax rate \eqref{eq:flat-rate} depends only on the player number $N$ and the budget factor $\bg$.

\begin{observation}
The efficient tax rate   \eqref{eq:flat-rate} increases with the player number $N$, and decreases with the budget factor $\bg$.
\end{observation}



\section{Fairness}\label{sec:framework-fairness}

We now study the taxing rule $\{\bex, \br\}$ that produces fair  equilibrium outcomes.
According to Theorems \ref{thm:efficiency} and   \ref{thrm:unique}, the flat tax rate in \eqref{eq:flat-rate} is sufficient for efficiency, while the tax exemptions $\bex$ do not affect efficiency.
This implies that we can adjust $\bex$,  while fixing the flat tax rate as in \eqref{eq:flat-rate}, to achieve the desired fairness without affecting the efficiency.

Substitute the flat tax rate \eqref{eq:flat-rate} into \eqref{eq:payoff-tax}, we can rewrite the payoff of each player $i$ in the Taxation Version $\G (\bex, \br) $ as:
\begin{equation}
 \ux_i(\bs) =  c \cdot \W(\bs) - c \cdot \Delta + \ex_i ,
\end{equation}
By Theorem \ref{thm:efficiency}, there   exists an efficient equilibrium $\bs^* $ of $\G (\bex, \br) $ that maximizes the social welfare of both $\G (\bex, \br) $ and $\G$, i.e., $\bs^* = \bs^\circ \eq \arg\max_{\bs} \W(\bs)$.
Under this efficient equilibrium $\bs^*$, each player's playoff is,
\begin{equation}\label{eq:payoff-exemption}
\ux_i(\bs^*) = c\cdot \overline{\W} - c \cdot \Delta+ \ex_i,
\end{equation}
where  $\overline{\W} \eq \W(\bs^\circ) $ is the maximum of  $\W(\bs)$.

By \eqref{eq:payoff-exemption}, we can see that each player's payoff, under the efficient equilibrium $\bs^*$ achieved from the flat tax rate  \eqref{eq:flat-rate}, is   determined by the tax exemptions $\bex $, as both $c$ and $\overline{\W}$ are constants and $\Delta$ depends on $\bex$ only.
Formally,
\begin{observation}\label{obs:exemption}
\emph{(i)} for any player $i\in \N$, his payoff $\ux_i(\bs^*)$ increases with $\ex_i$ and decreases with $\ex_j, \forall j\neq i$;
\emph{(ii)}
for any two players $i, j \in \N$,
$\ux_i(\bs^*) \geq \ux_j(\bs^*)$ if and only if $\ex_i \geq \ex_j$.
\end{observation}

Observation \ref{obs:exemption} implies that the tax exemptions $\bex $
can be used to adjust how the generated social welfare is distributed among players. More specifically,
\begin{itemize}[leftmargin=5mm]
  \item According to (i),  to increase (or decrease) the payoff of a player $i$, we can simply increase (or decrease) the tax exemption $\ex_i$ for player $i$;
  \item According to   (ii), to ensure   the payoff of player $i$ is larger  (or smaller)  than   player $j$, we need to set a larger  (or smaller) tax exemption $\ex_i$ for player $i$.
\end{itemize}
Based on the above, we can adjust  tax exemptions $\bex $ to achieve \emph{any desired fairness} together with   efficiency.

Next, we provide the tax exemptions $\bex $ for several typical fairness criterions such as the Max-min fairness \cite{max-min}, in which the social welfare is divided (among players) in such a way that the minimal payoff of players is maximized. 
Due to space limit, we put the detailed discussions regarding other fairness criterions in our online technical report \cite{report}.

For ease of illustration, we assume that $\bg = 1$ (strictly budget balance),
and hence,
$c = \frac{1}{N}$ and
$
\ux_i(\bs^*) = \frac{ \overline{\W}  }{N} -  \frac{  \Delta}{N}  + \ex_i
$.
Let $\omega \in (-\infty, \infty)$ denote an arbitrary real number.

\begin{lemma}[Max-min Fairness]\label{lemma:fair1}
If $  \ex_i =  \ex_j = \omega$,  $\forall i,j\in\N$ (i.e., flat tax exemption), then
$$
\textstyle
\ux_i(\bs^*) = \frac{ \overline{\W}  }{N} , \quad \forall i \in \N.
$$
That is, all players share the total social welfare equally, which is a special case of the Max-min fairness \cite{max-min}
\end{lemma}




\section{Implementation Issue}\label{sec:framework-implement}


%

To illustrate the implementation of the proposed taxation mechanism in practice, we now implement it in the Prisoner's Dilemma \cite{game-theory1994} as an illustrative example.

In the Prisoner's Dilemma game, two players make choices with strategic inter-dependence.
Each player can choose the strategy ``Cooperate (C)'' or ``Defect (D)''.
The payoffs of players under different strategy profiles are illustrated in
Table \ref{tbl:1}(a), where each row (or column) denotes the choice of player 1 (or 2), and the numbers in each cell denote the payoffs of both players under a particular strategy profile.
We can see that the unique NE of this game is (D, D), where both players choose ``Defect''.
However, the unique SE is (C, C), where both players   choose ``Cooperate''.

\begin{table} [t]
\centering
(a)\qquad\qquad\qquad\qquad\qquad\qquad(b)
\scalebox{0.9}[0.9]{
	\begin{tabular}{|c|c|c|}
	\hline
	           & C &	D
	 \\
	 \hline
	 C & $(2 ,\ 2 )$  & $(0 ,\ 3 ) $
	 \\
	 \hline
	 D & $(3 ,\ 0 )$ & $\mathbf{(1 ,\ 1 )}$
	 \\
	 \hline
	\end{tabular}
\qquad
    \begin{tabular}{|c|c|c|}
	\hline
	 & C &	D
	 \\
	 \hline
	 C	& $\mathbf{(2 ,\ 2  )}$ & $( {1.5},\  {1.5})$
	 \\
	 \hline
	 D &  $(  {1.5},\  {1.5})$ & $(1 ,\ 1 )$
	 \\
	 \hline
	\end{tabular}
}
\vspace{2mm}
\caption{(a) The Prisoner's Dilemma Game; (b) The Taxation Version  with $\r_1 = \r_2 = 0.5$ and $\ex_1= \ex_2 = 0$.}\label{tbl:1}
\vspace{-6mm}
\end{table}

%

Now we implement the Taxation Mechanism to the Prisoner's Dilemma game.
For ease of illustration, we assume that $\bg = 1$ (strictly budget balance).
According to Theorem \ref{thm:efficiency}, the efficient taxing rule is: $\r = \frac{N-1}{N} = 0.5$.
Note that the tax exemptions will not affect the efficiency, and thus we take the zero tax exemptions (i.e., $\ex_1= \ex_2 = 0$) as an example.
In this case, the new payoffs of players (after taxation) can be computed by \eqref{eq:xxx}, and illustrated in Table \ref{tbl:1}(b).
We can see that the unique NE  of the new Taxation Version game is (C, C), which is an efficient outcome of both  games.

\section{Conclusion}\label{sec:conclusion}

In this work, we propose a Taxation Mechanism and design the associated
taxing rule for inducing efficient and fair equilibrium in static strategic games.
This framework provides rich
modeling  possibilities, and spans the
continuum space between strategic interactions (of selfish players)
and altruistic interactions (of unselfish players).
We derive the tax rates which are necessary and sufficient
for achieving an efficient outcome in any static strategic game with common knowledge,
and provide several tax exemption examples that achieve some typical
fairness criterions. 

There are several interesting future directions.
First, the current Taxation Mechanism is only applicable for   games
with common knowledge.
It  would be interesting  to extend it to
games with incomplete information.
Second, in the current framework, we apply a single tax rate for each player.
It is meaningful to extend it to the case of progressive  tax rates as in real-world income tax systems.




\end{document}